\documentclass{iopart}
\usepackage[dvips]{graphicx}
\usepackage{iopams}
\usepackage{setstack}
\usepackage{multicol}
\usepackage{aeitensor}
\usepackage{multirow}
\usepackage{url}
\newcommand{\txt}[1]{\mathrm{#1}}
\newcommand{\abs}[1]{\left|#1\right|}
\newcommand{\un}[1]{\txt{\,#1}}
\newcommand{\mc}[1]{\mathcal{#1}}
\newcommand{\mf}[1]{\mathfrak{#1}}
\newcommand{\wt}[1]{\widetilde{#1}}
\newcommand{\F}{\mathcal{F}}
\newcommand{\Real}{\mathop{\rm Re}\nolimits}
\newcommand{\Imag}{\mathop{\rm Im}\nolimits}
\newcommand{\sinc}{\mathop{\rm sinc}\nolimits}

\newcommand{\khat}{\zhat}
\newcommand{\nhat}{\widehat{n}}

\newcommand{\zhat}{\widehat{k}}
\newcommand{\xihat}{\widehat{\xi}}
\newcommand{\etahat}{\widehat{\eta}}

\newcommand{\mbf}[1]{\ensuremath{\mathchoice{\mbox{\boldmath$\displaystyle#1$}}
{\mbox{\boldmath$\textstyle#1$}}
{\mbox{\boldmath$\scriptstyle#1$}}
{\mbox{\boldmath$\scriptscriptstyle#1$}}}}

\newcommand{\tens}[1]{\aeitensor{#1}}

\newcommand{\tauref}{\tau_{\txt{ref}}}
\newcommand{\synthLISA}{\txt{synthLISA}}
\newcommand{\LISAsim}{\txt{LISAsim}}

\newcommand{\SFT}{\txt{SFT}}
\newcommand{\LWL}{_\mathrm{LWL}}

\newcommand{\doppler}{\theta}
\newcommand{\ddoppler}{\Delta\theta}
\newcommand{\Fisher}{\bar{\Gamma}}

\newcommand{\sig}{{\mathrm{s}}}

\newcommand{\ampErr}{\epsilon_\Amp}
\newcommand{\dopErr}{\epsilon_\doppler}
\newcommand{\delA}{{\delta_\Amp}}
\newcommand{\phiA}{{\phi_\Amp}}
\newcommand{\cand}{{\mathrm{c}}}
\newcommand{\Akey}{\A_{\sig}}
\newcommand{\Acand}{\A_{\cand}}
\newcommand{\fkey}{f_{\sig}}
\newcommand{\fcand}{f_{\cand}}
\newcommand{\A}{\mc{A}}
\newcommand{\M}{\mc{M}}

\newcommand{\lvert}{|}
\newcommand{\rvert}{|}
\newcommand{\Amp}{\A}
\newcommand{\chalA}[1]{{#1}}
\newcommand{\chalB}[1]{{#1}}

\newcommand{\dcc}{LIGO-P080037-01-Z}
\newcommand{\Tlight}{T}

\newcommand{\coinc}{co\"{\i}ncidence}
\newcommand{\Coinc}{Co\"{\i}ncidence}
\newcommand{\coeff}{co\"{e}fficient}

\def\commitID{commitID: e2c3e2baa22c918ad9eb544aba2e7ae8dc6ccc94}
\def\commitDATE{ Fri Jun 27 15:50:17 2008 +0200}

\begin{document}
%\draft
\title[Improved $\F$-statistic search for white dwarf binaries in MLDC1B]
{Improved search for galactic white dwarf binaries\\
 in Mock LISA Data Challenge 1B\\ 
 using an $\F$-statistic template bank}
\address{\dcc\qquad\commitDATE \\\mbox{\small \commitID}}
\author{John T Whelan$^1$, Reinhard Prix$^2$ and Deepak Khurana$^3$}
\address{$^1$ Max-Planck-Institut f\"{u}r Gravitationsphysik
  (Albert-Einstein-Institut), D-14476 Potsdam, Germany}
\address{$^2$ Max-Planck-Institut f\"{u}r Gravitationsphysik
  (Albert-Einstein-Institut), D-30167 Hannover, Germany}
\address{$^3$ Indian Institute of Technology, Kharagpur, West Bengal
  721302, India}
\date{\commitDATE \\\mbox{\small \commitID}}
\begin{abstract}
  We report on our $\F$-statistic search for white-dwarf binary
  signals in the Mock LISA Data Challenge 1B (MLDC1B) .
  We focus in particular on the improvements in our search pipeline
  since MLDC1, namely refinements in the search pipeline and the use
  of a more accurate detector response (rigid adiabatic approximation).
  The search method employs a hierarchical template-grid based
  exploration of the parameter space, using a {\coinc} step to
  distinguish between primary (``true'') and secondary maxima,
  followed by a final (multi-TDI) ``zoom'' stage to provide an accurate
  parameter estimation of the final candidates.
\end{abstract}
\ead{John.Whelan@aei.mpg.de, Reinhard.Prix@aei.mpg.de}
% \maketitle

\section{Introduction}
\label{s:intro}

The Mock LISA Data Challenges (MLDCs)~\cite{mldc:_homepage} have the
purpose of encouraging the development of LISA data-analysis tools and
assessing the technical readiness of the community to perform
gravitational-wave (GW) astronomy with LISA.  The rounds so far have
been labelled MLDC1~\cite{Arnaud:2007vr}, MLDC2~\cite{Babak:2007zd}, and
MLDC1B~\cite{Babak:2008}.  The challenges have consisted of several
data-sets containing different types of simulated sources and LISA
noise. The three types of sources are white-dwarf binaries (WDBs),
co\"{a}lescing supermassive black holes (SMBHs) and extreme mass-ratio
inspirals (EMRIs).
GW signals from WDBs will be long-lasting
and \mbox{(quasi-)}monochromatic with
slowly-varying\footnote{In fact, the signals in the MLDCs so far have
  been strictly monochromatic.  Frequency evolution is being
  introduced for the first time in MLDC3.}
intrinsic frequency $f(\tau)$; in this sense they
belong to the class of \emph{continuous GWs}.
In the case of ground-based detectors the typical sources of
continuous GWs are spinning neutron stars with non-axisymmetric
deformations.  One of the standard tools developed for these searches
is the $\F$-statistic~\cite{Jaranowski:1998qm}, which corresponds to the
generalized log-likelihood ratio.

We have applied this method in our MLDC searches, adapting the
LAL/LALApps~\cite{lalapps} search code \texttt{ComputeFStatistic\_v2} used within the
LIGO Scientific Collaboration to search for periodic GW signals in
data from ground-based detectors such as LIGO and GEO\,600.
We have previously conducted searches for WDBs on data from
MLDC1~\cite{Prix:2007zh} and MLDC2~\cite{MLDC2Poster,MLDC2Paper}.
MLDC1B is a rerun of MLDC1 with different source parameters, and gives
us a chance to evaluate improvements in our pipeline since MLDC1.
Among the
issues encountered in our original MLDC1 analysis were inaccurate
determination of a subset of signal parameters due to the use of the
long-wavelength (LW) limit in modelling the LISA response, and the lack of
a method to distinguish secondary maxima in parameter space from
primary peaks of true signals.  Both of these aspects have been improved in our
MLDC1B pipeline.

\section{Continuous gravitational-wave signals}

A system with an oscillating mass quadrupole moment
emits GWs described, far from the source, by a metric perturbation
$\tens{h}$.
The WDB signals in the MLDCs 1, 2 and 1B have been restricted to
monochromatic signals with constant
intrinsic frequencies $f$, so in an inertial reference frame such as
the solar-system barycentre (SSB), the phase of this signal can be
written as $\phi(\tau) = 2\pi f (\tau - \tauref)$,
where $\tauref$ is a reference time.
We refer the reader to~\cite{Prix:2007zh} for a more
complete discussion of the formalism, here we only introduce the
notation and key results used in the following derivation.
The GW tensor can be expressed as
\begin{equation}
  \tens{h}(\tau) = \mc{A}^\mu\, \tens{h}_{\mu}(\tau) \,,
\end{equation}
where we introduce the convention of an implicit sum $\sum_{\mu=1}^4$ over
repeated indices $\mu, \nu$. The four \emph{amplitude parameters}
$\{\mc{A}^\mu\}$ are determined by the overall GW amplitude $h_0$,
the inclination angle $\iota$ of the orbital plane, the polarization
angle $\psi$, and the initial phase $\phi_0$. The explicit relations
$\mc{A}^\mu = \mc{A}^\mu(h_0, \iota, \psi, \phi_0)$ can be found in
Eq.(4) of~\cite{Prix:2007zh}.
The tensor wave components $\{\tens{h}_{\mu}\}$ depend on the
frequency $f$ and the propagation direction $\khat$ of the GW
(determined by the sky position of the source), namely
\begin{equation}
  \label{e:tensortemplates}
  \eqalign{
    \tens{h}_{1}(\tau; \doppler) = \tens{\varepsilon}_{\!+}(\khat)
    \, \cos\phi(\tau)
    \ ,
    &\qquad
    \tens{h}_{2}(\tau; \doppler) = \tens{\varepsilon}_{\!\times}(\khat)
    \, \cos\phi(\tau)
    \ ,
    \\
    \tens{h}_{3}(\tau; \doppler) = \tens{\varepsilon}_{\!+}(\khat)
    \, \sin\phi(\tau)
    \ ,
    &\qquad
    \tens{h}_{4}(\tau; \doppler) = \tens{\varepsilon}_{\!\times}(\khat)
    \, \sin\phi(\tau)
    \ ,
  }
\end{equation}
where we denote the set of \emph{Doppler parameters} $\doppler \equiv
\{f, \khat\}$. The polarization basis $\tens{\varepsilon}_{+,\times}$ associated
with the sky position is defined in terms of the right-handed
orthonormal basis
$\{\xihat,\,\etahat,\,\khat\}$ with
$\xihat$ lying in the ecliptic plane and $\etahat$ in the
northern hemisphere, namely
$\tens{\varepsilon}_{\!+} = \xihat \otimes \xihat - \etahat \otimes \etahat$
and
$\tens{\varepsilon}_{\!\times} = \xihat \otimes \etahat + \etahat \otimes \xihat$.

\section{The $\F$-statistic method}

The $\F$-statistic was originally developed in~\cite{Jaranowski:1998qm} and
extended to the multi-detector case in~\cite{Cutler:2005hc}.  A
generalization to the full TDI framework for LISA was developed
in~\cite{Krolak:2004xp}.  In this work we present an approach that
unifies the method for space- and ground-based detector data, allowing
for a more direct application of existing LIGO/GEO\,600 codes to LISA
data analysis.

A ``detector'' $I$ (here a TDI observable), provides a linear
transformation of the tensor metric perturbation
$\tens{h}_{\mu}(\tau)$ into a scalar ``signal'' $h^{I}_{\mu}(t)$ as a
function of detector time, so the detector output can be written as
\begin{equation}
  x^{I}(t) = n^{I}(t) + \mc{A}^\mu\, h^{I}_\mu(t; \doppler)\,,
\end{equation}
where $n^{I}(t)$ is the instrumental noise in detector $I$.
Following the notation of~\cite{Cutler:2005hc,Prix:2006wm}, we
write the different data-streams $x^I(t)$ as a vector $\mbf{x}(t)$,
and we define the standard multi-detector (with uncorrelated noise)
scalar product  as
\begin{equation}
  \label{e:innprod}
  (\mbf{x}|\mbf{y}) = \sum_{\alpha} \sum_{I} \int_{-\infty}^{\infty}
  \wt{x}_{\alpha}^{I*}(f)\, [S_{\alpha\,I}(f)]^{-1}
  \, \wt{y}_{\alpha}^I(f)\, df
  \ .
\end{equation}
Here we have broken up the observation time into intervals labelled by
$\alpha$, $\wt{x}_{\alpha}$ is the Fourier-transform of the data in
the $\alpha$th time interval, $x^*$ denotes complex conjugation, and
$\{S_{\alpha\,I}(f)\}$ is the two-sided noise power spectral
density appropriate to the $\alpha$th time interval. %%\footnote{We limit
%%  attention to the case of coherent combination of detectors with
%%  uncorrelated noise.}
%
We search for a signal $\{\Amp_\sig, \doppler_\sig\}$ by seeking the parameters
$\{\Amp_\cand, \doppler_\cand\}$
which maximize the log-likelihood ratio
\begin{equation}
  \hspace*{-1cm}
  L(\mbf{x}; \Amp, \doppler )
  = (\mbf{x}|\mbf{h}) - \frac{1}{2}(\mbf{h}|\mbf{h})
  = \mc{A}^\mu (\mbf{x}|\mbf{h}_\mu)
  - \frac{1}{2}\mc{A}^\mu (\mbf{h}_\mu|\mbf{h}_\nu) \mc{A}^\nu\,,
\end{equation}
with automatic summation over
repeated amplitude indices $\mu,\nu$. Defining
\begin{equation}
  x_\mu(\doppler) \equiv (\mbf{x}|\mbf{h}_\mu)\,,\quad\mbox{and}\quad
  \mc{M}_{\mu\nu}(\doppler) \equiv (\mbf{h}_\mu|\mbf{h}_\nu)\,,
  \label{e:Ametric}
\end{equation}
we see that $L$ is maximized for given $\doppler$ by the amplitude
estimator $\Acand^\mu = \mc{M}^{\mu\nu}x_\nu$, where
$\mc{M}^{\mu\nu}$
is the inverse matrix of $\mc{M}_{\mu\nu}$. Thus the detection
statistic $L$, maximized over the amplitude parameters $\Amp$, is
$\F$, where
\begin{equation}
  2\F(\mbf{x}; \doppler) \equiv \, x_\mu \, \mc{M}^{\mu\nu} \,x_\nu\,.
  \label{eq:2}
\end{equation}
This defines the (multi-detector) $\F$-statistic. One can show that
the expectation in the perfect-match case $\doppler = \doppler_\sig$
is $E[2\F(\doppler_\sig)] = 4 + \abs{\mc{A}_\sig}^2$, where we used the
definition
\begin{equation}
  \label{eq:4}
  \abs{\mc{A}}^2 \equiv \mc{A}^\mu \mc{M}_{\mu\nu} \mc{A}^\nu\,,
\end{equation}
for the norm of a 4-vector $\mc{A}^\mu$, using $\mc{M}_{\mu\nu}$ as a
\emph{metric} on the amplitude-parameter space.
Note that $\abs{\mc{A}_\sig}$ is the (optimal) signal-to-noise ratio
(SNR)  of the true signal $\{\mc{A}_\sig, \doppler_\sig\}$.

\section{Modelling the LISA response}

\label{s:resp}

The MLDC data were generated by two different programs: Synthetic
LISA~\cite{synthLISA} simulates a detector output consisting of Doppler
shifts of the LISA lasers due to relative motion of the spacecraft,
while LISA Simulator~\cite{LISAsim} simulates the phase differences
between laser light following different paths between the spacecraft.
In both cases the underlying variables are combined with appropriate
time shifts to form TDI observables which cancel the (otherwise
dominating) laser frequency
noise~\cite{TDI:_1999,Tinto:2003vj,Krolak:2004xp}.  One choice of such
TDI quantities is the set of three observables $\{X, Y, Z\}$, which
were used to publish the data of the first and second MLDCs.
These observables, which can be thought of as representing the output
of three virtual ``detectors'' (which we label with the index $I$),
are related to the gravitational wave tensor
$\tens{h}$ through the detector ``response'', which can be modelled
at different levels of accuracy. In the following we discuss two such
approximations for the response, the simple ``long-wavelength limit''
and the more accurate ``rigid adiabatic approximation''.

\subsection{Long-wavelength limit (LWL) response}
\label{sec:long-wavel-limit}

In the LWL approximation the reduced wavelength
$c/(2\pi f)$ is assumed to be large compared to the distance $L$
between the spacecraft, which corresponds to a light-travel time of
$\Tlight = L/c \sim 17\un{s}$ (assuming equal arm-lengths), and so this
approximation requires $f \ll 10\un{mHz}$.
In this approximation the GW contribution to each observable
can be modelled as
\numparts
  \label{eq:9}
  \begin{eqnarray}
    X^{\synthLISA} &\approx -{4 \Tlight^2} \,\tens{d}^{X}_{\LWL}
    : \frac{d^2\tens{h}}{dt^2}
    \ ,
    \\
    X^{\LISAsim} &\approx -{2 \Tlight} \, \tens{d}^{X}_{\LWL}
    : \frac{d\tens{h}}{dt}
    \ ,
  \end{eqnarray}
\endnumparts
where $:$ denotes the contraction of both tensor indices, and
$\tens{d}^{X}_{\LWL}\equiv(\nhat_2\otimes\nhat_2 -
\nhat_3\otimes\nhat_3)/2$ is the usual LWL response tensor for a GW
interferometer with arms $\nhat_2$ and $\nhat_3$. The analogous
expressions for $Y$ and $Z$ are obtained by cyclic permutations of the
indices $1\rightarrow 2\rightarrow 3\rightarrow 1$.  In the remainder
of this section we will give explicit expressions associated with the
$X$ variable, with the understanding that the formulas related to $Y$
and $Z$ can be constructed by analogy.

It is convenient to describe the ``response'' of a gravitational wave
detector in the frequency domain in terms of a response function
$R(f)$, relating the detector output to a ``strain'' more closely
connected to the metric perturbation tensor $\tens{h}$, so that
\begin{equation}
  \wt{X}(f) = \frac{\wt{h}^{X}(f)}{R(f)}
  = \frac{\tens{d}^{X} : \wt{\tens{h}}(f)}{R(f)}
\end{equation}
In the
long-wavelength limit, $\tens{d}^{X}\approx\tens{d}^{X}_{\LWL}$ and
\numparts
  \begin{eqnarray}
    R^{\synthLISA}(f) \approx R^{\synthLISA}_{\LWL}(f)
    &= \left(\frac{1}{4\pi f \Tlight}\right)^2
    \\
    R^{\LISAsim}(f) \approx R^{\LISAsim}_{\LWL}(f)
    &= i \frac{1}{4\pi f \Tlight}
    \ .
  \end{eqnarray}
\endnumparts
This formalism is valid in the regime where the finite lengths of data
used to approximate the idealized Fourier transforms are short enough
that the geometry and orientation of the detector doesn't change
significantly during this time.

\subsection{Rigid adiabatic approximation}
\label{sec:rigid-adiab-appr}

A more accurate approximation to the TDI response is the so-called
rigid adiabatic approximation~\cite{Rubbo:2003ap}, which for a wave
propagating along the unit vector $\khat$ results in
\begin{equation}
  \label{eq:RAA-response}
  \fl
  \frac{\tens{d}^X(f,\khat)}{R(f)} =
  \frac{e^{-i4\pi f\Tlight}}{R_{\LWL}(f)}
  \sinc\left({2\pi f\Tlight}\right)
  \left\{
    \mf{T}_{\nhat_2}(f,\khat)\frac{\nhat_2\otimes\nhat_2}{2}
    - \mf{T}_{-\nhat_3}(f,\khat)\frac{\nhat_3\otimes\nhat_3}{2}
  \right\}
\end{equation}
where (defining $\xi(\khat)\equiv 1-\khat\cdot\nhat$)
\begin{equation}
  \fl
  \mf{T}_{\nhat}(f,\khat)
  =  \frac{e^{i2\pi f\Tlight\khat\cdot\nhat/3}}{2}
\{
    e^{i\pi f\Tlight\xi(\khat)}
    \sinc
[
      \pi f\Tlight\xi(-\khat)
]
    +
    e^{-i\pi f\Tlight\xi(-\khat)}
    \sinc
[
      \pi f\Tlight\xi(\khat)
]
\}
\end{equation}
is a transfer function associated with the arm along $\nhat$.  Note
that this is related to the $\mc{T}_{\nhat}(f,\khat)$ defined
in~\cite{Rubbo:2003ap} by an overall phase, and also that
$\mf{T}_{\nhat}(f,\khat)$ reduces to unity in the LWL
$f\ll 1 / (\pi\Tlight)$.
For the separation of \eref{eq:RAA-response} into a response function
$R(f)$ and a detector tensor $\tens{d}^X(f,\khat)$, we choose
\begin{equation}
  \label{eq:RAA-d}
  \tens{d}^X(f,\khat) =
  \left\{
    \mf{T}_{\nhat_2}(f,\khat)\,\frac{\nhat_2\otimes\nhat_2}{2}
    - \mf{T}_{-\nhat_3}(f,\khat)\,\frac{\nhat_3\otimes\nhat_3}{2}
  \right\}
\end{equation}
\begin{equation}
  \label{eq:RAA-R}
  R(f) = \frac{ R_{\LWL}(f) \, e^{i4\pi f\Tlight}}{\sinc\left({2\pi f\Tlight}\right)}\,.
\end{equation}

\subsection{Calibrated SFTs}

The input to the LAL/LALApps search code consists of
Fourier-transformed data stretches of duration $T_\SFT$, referred to as
Short Fourier Transforms (SFTs).  This is a common data format used
within the LIGO Scientific Collaboration for continuous-wave searches
(e.g., see~\cite{Abbott:2006vg}).  The time baseline $T_\SFT$
has to be chosen sufficiently short such that the noise-floor can be
approximated as stationary and the rotation and acceleration of the
LISA detector can be neglected, and we chose $T_\SFT = 7$\,days.

We produce ``calibrated SFTs'' by Fourier-transforming the raw TDI data
and applying a
frequency-domain response function to produce a Fourier transformed
strain (including noise) of
\begin{equation}
  \wt{x}^{X}(f) \equiv R(f) \, \wt{X}(f)\,.
\end{equation}
For our MLDC1 analysis~\cite{Prix:2007zh} and MLDC2
submission~\cite{MLDC2Poster} we used the long-wavelength approximation
$R_{\LWL}(f)$ for calibrating SFTs, but for subsequent
analyses we have produced ``rigid adiabatic'' SFTs, which use the
full form of $R(f)$ defined in \eref{eq:RAA-R}.

\subsection{Modelling of detector response in different analyses}
\label{ss:resp-comp}

Our MLDC1B pipeline includes modifications to implement the full form of
$\tens{d}^{I}(f,\khat)$.  However, a logistically simpler
intermediate approximation was also used in the initial followup to our
MLDC2 work.  In this ``partial rigid adiabatic'' formalism, the more
precise form of $R(f)$ from \eref{eq:RAA-R} is used to
construct the SFTs, but the further analysis proceeds with the simpler
form of $\tens{d}^{I}_{\LWL}$. See \tref{tab:RAdefns} for a summary of the
three different levels of response approximation considered in this analysis.
\begin{table}[tbp]
  \caption{Definitions of the long-wavelength (LW), partial rigid
    adiabatic (pR), and full rigid adiabatic (RA) formalisms, in terms
    of the response function $R(f)$ (used to
    calibrate SFTs) and the detector tensor $\tens{d}^{I}(f,\khat)$.}
  \label{tab:RAdefns}

  \begin{indented}
  \item[]\begin{tabular}{@{}l ccc}
      \br
      full name & label & response & detector tensor \\
      \mr
      long-wavelength & LW & $R_{\LWL}(f)$ & $\tens{d}^{I}_{\LWL}$ \\
      partial rigid adiabatic & pR & $R(f)$ & $\tens{d}^{I}_{\LWL}$ \\
      full rigid adiabatic & RA & $R(f)$ & $\tens{d}^{I}(f,\khat)$ \\
      \br
    \end{tabular}
  \end{indented}
\end{table}

\section{Signal templates in the rigid adiabatic formalism}

\subsection{Amplitude Modulation {\coeff}s}

In the long-wavelength and \emph{partial} rigid adiabatic approximations, the
strain $h^{I}(t)$ at the detector is taken to be the contraction of the
metric perturbation $\tens{h}(t(\tau))$ with a detector tensor
$\tens{d}^{I}_{\LWL}(t)$ which is independent of the frequency and sky direction
of the signal, but which varies slowly with time due to the change of
orientation of the detector, in this case as LISA orbits the sun:
$h^{I}(t) = \tens{d}^{I}_{\LWL}(t):\tens{h}(t(\tau))$.
The template ``basis functions'' \eref{e:tensortemplates} therefore
read as
\begin{equation}
  \fl
  \eqalign{
    h^{I}_1(t;\doppler)
    = a_{\LWL}^{I}(t,\khat)\, \cos \phi\left(\tau(t;\doppler)\right),
    &\qquad
    h^{I}_2(t;\doppler)
    = b_{\LWL}^{I}(t,\khat)\, \cos\phi\left(\tau(t;\doppler)\right),
    \\
    h^{I}_3(t;\doppler)
    = a_{\LWL}^{I}(t,\khat)\, \sin\phi\left(\tau(t;\doppler)\right),
    &\qquad
    h^{I}_4(t;\doppler)
    = b_{\LWL}^{I}(t,\khat)\, \sin\phi\left(\tau(t;\doppler)\right)\,,
    }\label{eq:8}
\end{equation}
where we have defined the usual amplitude-modulation factors
$a_{\LWL}^{I}(t,\khat) \equiv
\tens{d}_{\LWL}^{I}(t):\tens{\varepsilon}_{\!+}(\khat)$ and
$b_{\LWL}^{I}(t,\khat) \equiv
\tens{d}_{\LWL}^{I}(t):\tens{\varepsilon}_{\!\times}(\khat)$.  In the
full rigid adiabatic analysis, however, we need to perform this
conversion in the frequency domain, because the response tensor
$\tens{d}^{I}(f,\khat)$ depends on the frequency of the incoming waves.
In the rigid adiabatic approximation, we model the slow time
dependence of the orientation of the detector by using a detector
tensor $\tens{d}_{\alpha}^{I}(f,\khat)$ appropriate for the time of
the $\alpha$th SFT.  This yields the templates
\begin{equation}
  \fl
  \eqalign{
    h^{I}_{\alpha,1}(f;\doppler)
    = a^{I}_\alpha(f,\khat)\,\wt{\cos\phi_\alpha}(f;\doppler)
    \ ,
    &\qquad
    h^{I}_{\alpha,2}(f;\doppler)
    = b^{I}_\alpha(f,\khat)\,\wt{\cos\phi_\alpha}(f;\doppler)
    \ ,
    \\
    h^{I}_{\alpha,3}(f;\doppler)
    = a^{I}_\alpha(f,\khat)\,\wt{\sin\phi_\alpha}(f;\doppler)
    \ ,
    &\qquad
    h^{I}_{\alpha,4}(f;\doppler)
    = b^{I}_\alpha(f,\khat)\,\wt{\sin\phi_\alpha}(f;\doppler)\,.
  }
  \label{e:freqtemplates}
\end{equation}
Now the amplitude modulation (AM) {\coeff}s are constructed from the
response tensor according to
\begin{equation}
    a^{I}_\alpha(f,\khat) = \tens{d}^{I}_\alpha(f,\khat)
    : \tens{\varepsilon}_{\!+}(\khat)
    \,,
    \qquad
    b^{I}_\alpha(f,\khat) = \tens{d}^{I}_\alpha(f,\khat)
    : \tens{\varepsilon}_{\!\times}(\khat)
    \ .
\end{equation}
Note that
the amplitude-modulation functions $a$ and $b$ now depend on
  frequency as well as sky position, contrary to the LWL case,
and they are now complex, due to the complex response tensor
\eref{eq:RAA-response}.

\subsection{Amplitude metric}

Using the rigid adiabatic forms of the templates
\eref{e:freqtemplates} gives the following form for the amplitude
metric $\M_{\mu\nu}$
defined in \eref{e:Ametric}:
\begin{equation}
  \label{eq:3}
  \{\mc{M}_{\mu\nu}\}
  =
  \left(
  \begin{array}{cccc}
    A &  C &  0 & E \\
    C &  B & -E & 0 \\
    0 & -E &  A & C \\
    E &  0 &  C & B
  \end{array}
  \right)
  \ .
\end{equation}
This form was first exhibited in~\cite{Krolak:2004xp}.  In our
notation, the non-zero amplitude metric elements for a signal with intrinsic
frequency $f_0$ are
\begin{equation}
  \fl
  \eqalign{
    A = \sum_\alpha\sum_I \frac{T_{\SFT}}{2S^I_\alpha(f_0)}
    \abs{a^I_{\alpha}(f_0)}^2
    \ ,
    &\qquad
    C = \sum_\alpha\sum_I \frac{T_{\SFT}}{2S^I_\alpha(f_0)}
    \Real[a^I_{\alpha}(f_0)^*b^I_{\alpha}(f_0)]
    \ ,
    \\
    B = \sum_\alpha\sum_I \frac{T_{\SFT}}{2S^I_\alpha(f_0)}
    \abs{b^I_{\alpha}(f_0)}^2
    \ ,
    &\qquad
    E = \sum_\alpha\sum_I \frac{T_{\SFT}}{2S^I_\alpha(f_0)}
    \Imag[a^I_{\alpha}(f_0)^*b^I_{\alpha}(f_0)]
  }
\end{equation}
Note that the off-block-diagonal element $E$ vanishes in the
long-wavelength (and partial rigid adiabatic) case because the
detector tensor $\tens{d}^{I}_{\LWL}$, and thus the AM {\coeff}s, are
real.

For this project, we have enhanced the $\F$-statistic search in the
LAL and LALApps libraries to allow for AM {\coeff}s which depend on
frequency and sky direction, and for the more complicated form of the
amplitude metric \eref{eq:3} arising from the complex AM {\coeff}s.

\section{Results and evaluation}

As in the first MLDC, the WDB portion of MLDC1B consisted of seven
challenges, labelled \mbox{1.1.1a-c} and \mbox{1.1.2-5}.  For each challenge,
``training'' and ``blind challenge'' datasets were provided, generated
with different randomly chosen sources.
We performed the
analysis for our challenge entry on the LISA Simulator data, but
discovered subsequently that inconsistent metadata in this dataset led
to a 7.5-second time offset and associated systematic error in the
initial phase.  We are therefore presenting here the results of an
analysis with the same pipeline on Synthetic LISA data.

\subsection{MLDC1B pipeline}

\begin{figure}[tbp]
  \centering
  \includegraphics[width=0.7\textwidth]{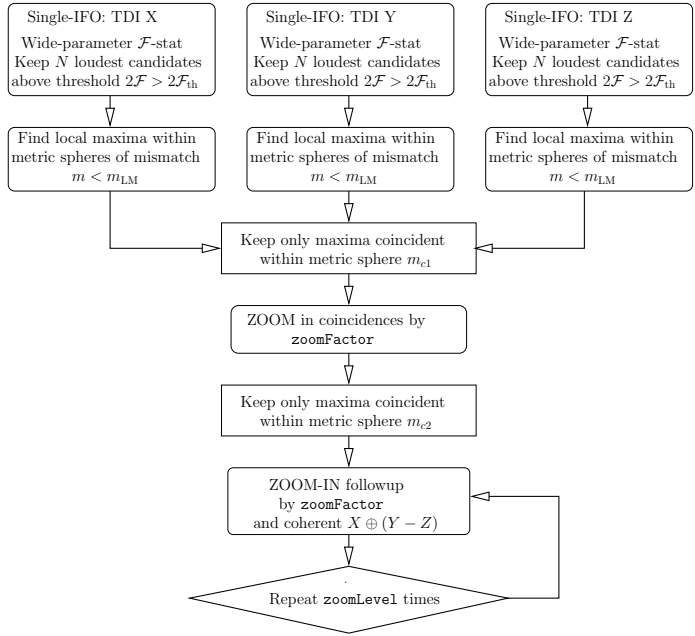}
  \caption{The pipeline used for MLDC1B. The pipeline settings used in
    this search were: initial template-grid mismatch $m_0 = 0.25$,
    local-maximum sphere of mismatch $m_{\mathrm{LM}} = 4.0$, first
    {\coinc}-stage mismatch $m_{c1} = 0.8$, followup {\coinc}
    mismatch $m_{c2} = 0.35$. The zoom-stage used \texttt{zoomLevels}
    = 2 and \texttt{zoomFactor}=10.
  }
  \label{f:pipeline}
\end{figure}

A major limitation of our MLDC1 analysis~\cite{Prix:2007zh} was the
lack of a robust method for distinguishing secondary maxima in Doppler
parameter space from additional true signals with lower $\F$-statistic
values.  Starting with our MLDC2 analysis~\cite{MLDC2Poster}, we
implemented a {\coinc} condition, where consistent signals are
required in searches performed will all three TDI variables ($X$, $Y$,
and $Z$).  This is described in detail in a forthcoming
paper~\cite{MLDC2Paper}.  \Fref{f:pipeline} illustrates the pipeline
used for this analysis.
The first stage consisted of wide-parameter searches over the Doppler
space $\doppler = \{f, \khat\}\,$, using each of the three TDI
variables $X$, $Y$ and $Z$ independently. The template bank used in this
first pipeline stage used an isotropic grid on the sky, with angular
mesh size $d \alpha = \sqrt{2 m_0} / (2\pi\,f\,R_\mathrm{orb}/c)$,
with the orbital radius $R_\mathrm{orb} = 1\,$AU.
The frequency spacing used was $d f = \sqrt{12 \,m_0} / ( \pi \, T)$,
where $T = 1\,$y is the observation time. The mismatch $m_0$ used in this
first stage template bank was $m_0 = 0.25$. The expressions for these step sizes were
computed from the orbital metric $g_{ij}$ \cite{Prix:2006wm}.
The mismatches used in the local-maxima finder and {\coinc} steps
were computed from this metric, using the definition
$m \equiv g_{ij} \Delta\doppler^i \Delta\doppler^j$, where $\Delta\doppler^i$ are
the Doppler coordinate differences between two candidates. Local
maxima are defined as the loudest candidate within a metric sphere of
mismatch $m < m_{\mathrm{LM}}$. {\Coinc} between $X$, $Y$ and $Z$
is defined a having at least one local maximum from each TDI variable
within a {\coinc}-window of mismatch $m < m_{c1}$ in the first {\coinc}
step, and $m_{c2}$ in the second (zoomed) {\coinc} step.
This {\coinc} scheme was found to be effective in eliminating
candidates related to secondary maxima.
Zooming of candidates was achieved by running a search covering 4
neighbouring template points in each dimension with a template-grid
resolution increased by a factor \texttt{zoomFactor},
i.e.\ $m_0' = m_0 / \mbox{\texttt{zoomFactor}}$.
The final zoom-stage serves only to increase the parameter-estimation
accuracy of the final candidates, and is using a coherent TDI combination
of the noise-independent variables $X$ and $Y-Z$.

\subsection{Evaluation}

\label{s:results-eval}

In order to evaluate our errors in parameter estimation, we compare
our estimates to the injected parameters (provided in the 'key', and
denoted by a subscript 's').
In Doppler space, $\Delta f = \fcand - \fkey$ denotes the frequency error,
and $\phi_{\txt{sky}}$ is the angle between recovered and true sky
position; they can be combined into $\dopErr^2 \equiv
\frac{1}{3}m\abs{\Akey}^2$, where $\dopErr$ measures the number of
``sigmas'' error in the Doppler mismatch between the
recovered parameters and the injected ones. This is based on
the Fisher-matrix $\Fisher_{ij}$, which is related to the Doppler
metric $g_{ij}$ via $\Fisher_{ij} = g_{ij}\,\abs{\Akey}^2$,
e.g., see~\cite{Prix:2006wm}, and where $m = g_{ij} \ddoppler^i \ddoppler^j$,
with $\ddoppler^i \equiv \doppler^i_\cand - \doppler^i_\sig$.
This error measure is chosen in such a way that in Gaussian noise
it should satisfy $E[\dopErr^2]=1$.
The Doppler metric $g_{ij}$ used here is a simplified ``phase
metric''; ideally the full $\F$-statistic metric should be used
instead~\cite{Prix:2006wm}, so this should only be considered a
rough estimate of the statistical significance of the Doppler errors
$\ddoppler$.
For challenges with multiple signals, the Doppler mismatch was also
used to distinguish found signals (where the candidate parameters were
within $m\le 1$ of the true signal) from false alarms (candidates
having no true signal within $m\le1$).

The
amplitude parameter errors are characterized by
$\delA \equiv \frac{\abs{\Acand}^2 - \abs{\Akey}^2 - 4}{2 \,\abs{\Akey}^2}$,
which measures the
error in the length of the 4D amplitude parameter vector, and
$\phiA$, which is the angle between the recovered and true
amplitude parameter vectors
(calculated using the amplitude metric $\mc{M}_{\mu\nu}(\doppler_\sig)$,
as in~\cite{Prix:2007zh}).
These two measures of amplitude-error are defined so that for statistical
errors due to Gaussian noise, the expectation is $E[\delA] = 0$
and the standard deviation of both $\delA$ and $\phiA$
is $\abs{\Akey}^{-1}$. The
amplitude parameter errors can be combined into a ``sigma'' error,
which is $\ampErr = \abs{\Delta\Amp}/2$, with $\Delta\Amp \equiv
\Acand - \Akey$, defined so that $E[\ampErr^2]=1$.
(Note that the use of $\delA$ for the error in the magnitude and
$\ampErr$ for the overall amplitude-error is a change in notation
from~\cite{Prix:2007zh}.)

\subsection{Isolated binaries (Challenge 1.1.1)}

\label{s:results-111}

\begin{figure}[tbp]
  \centering
  \includegraphics[width=\textwidth]{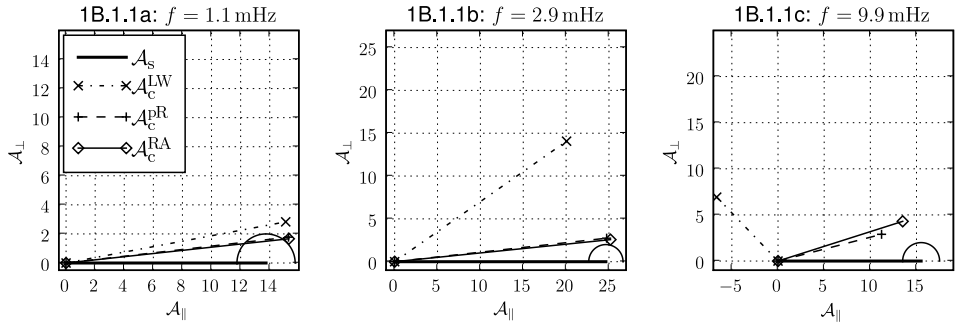}
  \caption{Recovery of amplitude parameters in Challenges 1B.1.1a
    (\emph{left}), 1B.1.1b (\emph{middle}), and 1B.1.1c
    (\emph{right}), using LW, pR and RA response models (see
    \tref{tab:RAdefns}). Each plot compares the recovered amplitude
    4-vector $\Acand$ to the injected signal 4-vector $\Akey$,
    shown in the plane defined by the two vectors.  Gaussian
    fluctuations would lead to a separation of the endpoints of the
    order $\abs{\Delta\Amp}\equiv\ampErr\sim 2$.
    (Note that the two-dimensional projection is
    slightly misleading:
    since all vectors do not lie in the same plane, the recovered
    amplitude vectors should be compared only with the key and not
    with each other.)
  }
  \label{f:1B_1_1-ampErrors}
\end{figure}

This challenge consisted of three separate data sets, each containing
one WDB signal at an unspecified sky position and within a given
frequency band: in 1.1.1a at $\sim 1\un{mHz}$, in 1.1.1b at $\sim
3\un{mHz}$, and in 1.1.1c at $\sim 10\un{mHz}$.  In each case we were able
to recover the injected signal, regardless of the model used for the
LISA response, but the recovered amplitude parameters were
considerably better, especially at high frequencies, using the full
rigid adiabatic response than the long-wavelength limit.  The recovery
of amplitude and phase parameters is illustrated in
\fref{f:1B_1_1-ampErrors} and \tref{tab:1B_1_1-errors}.

\begin{table}[tbp]
  \caption{
Recovery of Doppler- and amplitude parameters in Challenges \mbox{1B.1.1a-c}.
The results indicate adequate Doppler-parameter recovery regardless of
the response model used.
As seen previously in
\fref{f:1B_1_1-ampErrors}, recovery of \emph{amplitude} parameters improves
using the
full RA response, especially at higher frequencies.
Note, however, that the errors $\ampErr$ in amplitude parameters
with the full RA response still appear to be a bit larger
than the statistical expectation.
  }
  \label{tab:1B_1_1-errors}
  \begin{indented}
  \item[]\begin{tabular}{@{}lr ccc ccc}
\br
Challenge & Resp & $\Delta f$(nHz) & $\phi_{\txt{sky}}$(mrad) & $\dopErr$ & $\delA$ & $\phiA$ & $\ampErr$\\
\mr
1B.1.1a& RA & {$-0.68$} & {$46.06$} & {$0.54$} & {$0.12$} & {$0.11$} & {$1.14$} \\
($f=1.1$mHz; & pRA & {$-0.68$} & {$61.88$} & {$0.70$} & {$0.12$} & {$0.12$} & {$1.18$} \\
$\abs{\Akey}^{-1}=0.07$) & LW & {$-0.68$} & {$61.88$} & {$0.70$} & {$0.11$} & {$0.19$} & {$1.57$} \\
\br
1B.1.1b& RA & {$0.95$} & {$7.71$} & {$0.94$} & {$0.02$} & {$0.10$} & {$1.32$} \\
($f=2.9$mHz; & pRA & {$0.95$} & {$12.30$} & {$1.03$} & {$0.01$} & {$0.11$} & {$1.37$} \\
$\abs{\Akey}^{-1}=0.04$) & LW & {$0.95$} & {$12.30$} & {$1.03$} & {$-0.01$} & {$0.61$} & {$7.40$} \\
\br
1B.1.1c& RA & {$1.84$} & {$7.49$} & {$0.72$} & {$-0.09$} & {$0.31$} & {$2.37$} \\
($f=9.9$mHz; & pRA & {$1.84$} & {$5.12$} & {$0.67$} & {$-0.23$} & {$0.25$} & {$2.59$} \\
$\abs{\Akey}^{-1}=0.06$) & LW & {$1.84$} & {$5.12$} & {$0.67$} & {$-0.32$} & {$2.34$} & {$11.60$} \\
\br
\end{tabular}

  \end{indented}
\end{table}

\subsection{Verification binaries (Challenge 1.1.2)}

\begin{figure}[tbp]
  \centering
  \includegraphics[width=\textwidth]{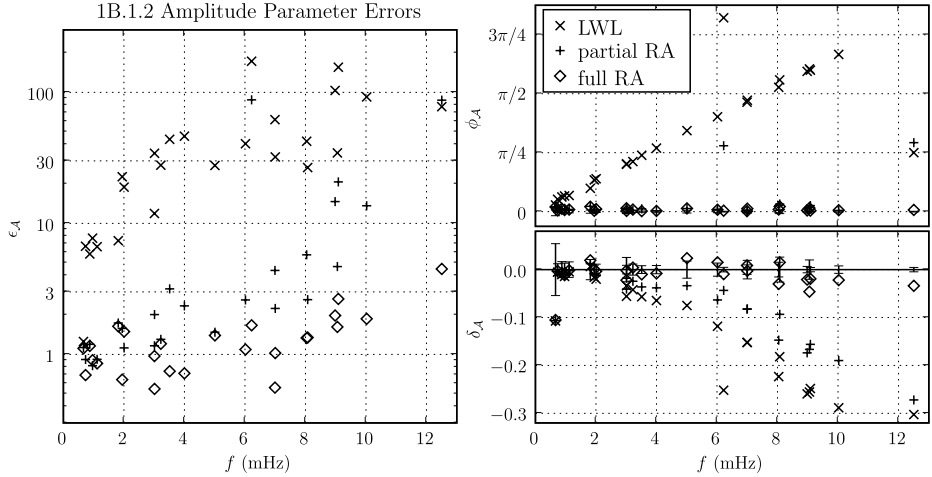}
  \caption{Amplitude parameter errors in Challenge 1B.1.2.  The
    quantities plotted are defined in \sref{s:results-eval}.  We
    see a continuation of the trends observed for isolated binaries:
    the LW response has phase errors ($\phiA$) which grow with
    frequency; at higher frequencies, both LW and partial RA exhibit
    a loss of SNR (negative $\delA$), which is much less pronounced
    with the full RA response. The latter nonetheless displays a
    slight systematic excess in overall error, i.e.~$\ampErr\gtrsim 1$.}
  \label{f:1B_1_2-ampErrors}
\end{figure}

In Challenge 1.1.2, the sky position and frequency of 25
``verification binaries'' was given, while the amplitude parameters of
the injected signals were unknown.  We therefore performed a targeted
$\mc{F}$-statistic search at each of the specified sets of Doppler
parameters, and found the maximum-likelihood estimators $\Acand$ for
the amplitude parameters.  These results are shown in
\fref{f:1B_1_2-ampErrors} and illustrate the performance of the
various LISA response models as seen already in Challenge 1.1.1.

\subsection{Recovery of multiple signals (Challenges 1.1.3-5)}

\begin{figure}[tbp]
  \centering
  \includegraphics[width=\textwidth]{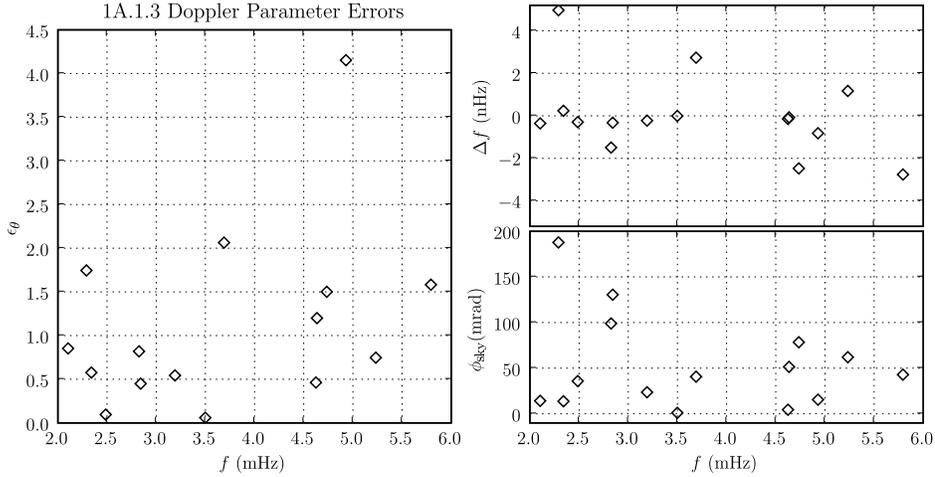}
  \caption{ Doppler parameter recovery in 1A.1.3.  The quantities
    plotted are defined in \sref{s:results-eval}.  Most signals
    were recovered with errors compatible with statistical
    expectations ($\dopErr\sim 1$); the few outliers may be due to
    source confusion.
  }
  \label{f:1A_1_3-dopErrors}
\end{figure}

Challenges 1B.1.3-5 were supposed to contain multiple binaries,
increasingly crowded in Doppler space.  Unfortunately, 1B.1.3 was
generated with no detectable signals; even the loudest had
$\abs{\Akey}^2\approx0.6$.  To check performance for resolvable,
detectable binaries, we consider the application of our MLDC1B search
pipeline both to MLDC1B data, and to the original MLDC1 datasets.  The
latter search we refer to MLDC1A, to distinguish it from the MLDC1
search we reported in~\cite{Prix:2007zh}.  The results are
summarized in \tref{tab:overTable}.  The corresponding Doppler
parameter recovery for 1A.1.3 is shown in \fref{f:1A_1_3-dopErrors}.
\begin{table}[tbp]
  \caption{
    Summary of the number of signals found, signals missed, and false alarms in Challenges
    \chalA{1A.1.3-5} and \chalB{1B.1.3-5}. We distinguish missed signals
    with $\abs{\Akey}^2>40$, which should in principle be
    detectable with our current pipeline, and those with
    $\abs{\Akey}^2<40$, which are likely to be too weak to pass our
    detection threshold.  Challenge 1B.1.3 contained no detectable signal,
    and the only candidate returned by our search was one
    low-significance false alarm with $2\F \sim 30$.
  }
  \label{tab:overTable}
  \begin{indented}
  \item[]\begin{tabular}{@{}l c rl rl rl rl}
\br
&  & \multicolumn{2}{c}{Found} & \multicolumn{4}{c}{Missed} & \multicolumn{2}{c}{False} \\
&  &  &  & \multicolumn{2}{r}{$\lvert\Akey\rvert^2 > 40$} & \multicolumn{2}{l}{$\lvert\Akey\rvert^2 < 40$} &  \\
Chal & $f$ range (mHz) & 1A & 1B & 1A & 1B & 1A & 1B & 1A & 1B \\
\mr
1.1.3 & $2-7$ & \chalA{15} & \chalB{0} & \chalA{5} & \chalB{0} & \chalA{0} & \chalB{20} & \chalA{1} & \chalB{1} \\
1.1.4 & $3.000-3.015$ & \chalA{15} & \chalB{13} & \chalA{27} & \chalB{22} & \chalA{3} & \chalB{17} & \chalA{5} & \chalB{1} \\
1.1.5 & $2.9985-3.0015$ & \chalA{3} & \chalB{3} & \chalA{30} & \chalB{29} & \chalA{0} & \chalB{12} & \chalA{0} & \chalB{0} \\
\br
\end{tabular}

  \end{indented}
\end{table}
Note that our signal recovery in challenges 1.1.4 and 1.1.5 is still
limited by source confusion, where the primary maximum of a weaker
signal can be modified substantially by interference with the
secondary peak-structure of stronger signals.

\section{Conclusions and outlook}

We implemented an $\F$-statistic search for white-dwarf binaries in
the Mock LISA Data Challenge based on the LAL/LALApps code
developed to search for spinning neutron stars in LIGO/GEO\,600 data.
MLDC1B has given us the opportunity to improve our MLDC1 search,
replacing the long-wavelength response with the
rigid adiabatic formalism more appropriate to LISA data analysis.
We see that our amplitude parameter recovery is
markedly improved by this enhancement, especially at frequencies above
$5\un{mHz}$.

Another difference from our MLDC1 search is an improved
discrimination between secondary maxima and true signals, which is
accomplished using a {\coinc} condition between searches on different
TDI variables.  This is most relevant in the case of a large
number of signals, such as in MLDC2, and a separate paper will address our
findings in detail~\cite{MLDC2Paper}.

More work remains to improve our handling of source confusion and deal
with multiple interfering sources, as illustrated by the limited
number of sources recovered in challenges 1.1.4 and 1.1.5.

\ack
This work was supported by the Max-Planck-Society and the German
Aerospace Center (DLR).
DK would like to thank the Albert Einstein Institute (Max Planck
Institute for Gravitational Physics) for support and hospitality.
This paper has been assigned LIGO Document Number \dcc.

\section*{References}
\bibliography{biblio}

\providecommand{\newblock}{}
\begin{thebibliography}{10}
\expandafter\ifx\csname url\endcsname\relax
  \def\url#1{{\tt #1}}\fi
\expandafter\ifx\csname urlprefix\endcsname\relax\def\urlprefix{URL }\fi
\providecommand{\eprint}[2][]{\url{#2}}
% Bibliography created with iopart-num v2.0
% /biblio/bibtex/contrib/iopart-num

\bibitem{mldc:_homepage}
{MLDC} homepage \url{http://astrogravs.nasa.gov/docs/mldc/}

\bibitem{Arnaud:2007vr}
Arnaud K~A {\em et~al.\/} 2007 {\em Class. Quant. Grav.\/} {\bf 24} S529--S540
  (\textit{Preprint} \eprint{gr-qc/0701139})

\bibitem{Babak:2007zd}
Babak S {\em et~al.\/} (Mock LISA Data Challenge Task Force) 2007
  (\textit{Preprint} \eprint{0711.2667})

\bibitem{Babak:2008}
Babak S {\em et~al.\/} (Mock LISA Data Challenge Task Force) 2008

\bibitem{Jaranowski:1998qm}
Jaranowski P, Krolak A and Schutz B~F 1998 {\em Phys. Rev.\/} {\bf D58} 063001
  (\textit{Preprint} \eprint{gr-qc/9804014})

\bibitem{lalapps}
{LIGO Scientific Collaboration} {LAL/LALApps: FreeSoftware (GPL) tools for
  data-analysis.} \url{http://www.lsc-group.phys.uwm.edu/daswg/}

\bibitem{Prix:2007zh}
Prix R and Whelan J~T 2007 {\em Class. Quant. Grav.\/} {\bf 24} S565--S574
  (\textit{Preprint} \eprint{0707.0128})

\bibitem{MLDC2Poster}
Prix R and Whelan J~T 2007 {F-Statistic Search on the Second Mock LISA Data
  Challenge} LIGO-G070462-00-Z poster at VIIIth Edoardo Amaldi Conference

\bibitem{MLDC2Paper}
Whelan J~T, Prix R and Khurana D 2008

\bibitem{Cutler:2005hc}
Cutler C and Schutz B~F 2005 {\em Phys. Rev.\/} {\bf D72} 063006
  (\textit{Preprint} \eprint{gr-qc/0504011})

\bibitem{Krolak:2004xp}
Krolak A, Tinto M and Vallisneri M 2004 {\em Phys. Rev.\/} {\bf D70} 022003
  (\textit{Preprint} \eprint{gr-qc/0401108})

\bibitem{Prix:2006wm}
Prix R 2007 {\em Phys. Rev.\/} {\bf D75} 023004 (\textit{Preprint}
  \eprint{gr-qc/0606088})

\bibitem{synthLISA}
Vallisneri M {Synthetic LISA Software}
  \url{http://www.vallis.org/syntheticlisa/}

\bibitem{LISAsim}
{Cornish} N~J and Rubbo L {The LISA Simulator}
  \url{http://www.physics.montana.edu/lisa/}

\bibitem{TDI:_1999}
Armstrong J, Estabrook F and Tinto M 1999 {\em The Astrophysical Journal\/}
  {\bf 527} 814--826

\bibitem{Tinto:2003vj}
Tinto M, Estabrook F~B and Armstrong J~W 2004 {\em Phys. Rev.\/} {\bf D69}
  082001 (\textit{Preprint} \eprint{gr-qc/0310017})

\bibitem{Rubbo:2003ap}
Rubbo L~J, Cornish N~J and Poujade O 2004 {\em Phys. Rev.\/} {\bf D69} 082003
  (\textit{Preprint} \eprint{gr-qc/0311069})

\bibitem{Abbott:2006vg}
Abbott B {\em et~al.\/} (LIGO Scientific) 2007 {\em Phys. Rev.\/} {\bf D76}
  082001 (\textit{Preprint} \eprint{gr-qc/0605028})

\end{thebibliography}

\end{document}